\documentclass{aastex62}

\usepackage[T1]{fontenc}
\usepackage{ae,aecompl}

\usepackage{graphicx}	
\usepackage{amsmath}	
\usepackage{amssymb}	
\usepackage{soul}		
\usepackage{natbib}
\usepackage{color}
\definecolor{red}{rgb}{1,0,0}
\definecolor{blue}{rgb}{0,0,1}
\definecolor{green}{rgb}{0,1,0}

\received{2019 December 16}
\accepted{2020 January 2}
\published{2020 January 3}

\shorttitle{GW170729's family tree}
\shortauthors{Kimball et al.}

\begin{document}

\title{What GW170729's exceptional mass and spin tells us about its family tree}

\correspondingauthor{Chase~Kimball}
\email{CharlesKimball2022@u.northwestern.edu}

\author[0000-0001-9879-6884]{Chase~Kimball}

\author[0000-0003-3870-7215]{Christopher~P~L~Berry}

\author[0000-0001-9236-5469]{Vicky~Kalogera}

\affiliation{Center for Interdisciplinary Exploration and Research in Astrophysics (CIERA), Department of Physics and Astronomy, Northwestern University, 1800 Sherman Ave, Evanston, IL 60201, USA}

\begin{abstract}
Gravitational-wave observations give a unique insight into the formation and evolution of binary black holes. We use gravitational-wave measurements to address the question of whether GW170729's source, which is (probably) the most massive binary and the system with the highest effective inspiral spin, could contain a black hole which is a previous merger remnant. Using the inferred mass and spin of the system, and the empirically determined population of binary black holes, we compute the evidence for the binary being second-generation compared with first-generation. We find moderate evidence (a Bayes factor of $\sim6$--$7$) that the mass and spin better match a second-generation merger, but folding in the expectation that only a small fraction of mergers are second-generation, we conclude that there is no strong evidence that GW170729 was the result of a second-generation merger. The results are sensitive to the assumed mass distribution, and future detections will provide more robust reconstructions of the binary black hole population.
\end{abstract}


\section{Introduction}

LIGO--Virgo \citep{TheLIGOScientific:2014jea,TheVirgo:2014hva} have observed gravitational waves from $10$ binary black holes \citep{LIGOScientific:2018mvr}. 
Of these, GW170729's source stands out as (probably) the system with the highest mass and the highest effective inspiral spin $\chi_\mathrm{eff}$. With a total mass of $M = 84.4^{+15.8}_{-11.1}M_{\odot}$ and a chirp-mass of $\mathcal{M} = 35.4^{+6.5}_{-4.8}M_\odot$, the primary-component mass $m_1 = 50.2^{+16.2}_{-10.2}M_{\odot}$ 
encroaches on the hypothesised (pulsational) pair-instability supernovae mass-gap \citep{Woosley:2016hmi}. 
Its $\chi_\mathrm{eff} = 0.37^{+0.21}_{-0.25}$ makes it one of two observations for which a non-spinning component is excluded at $90\%$ probability. 
Given GW170729's exceptional properties, it is natural to ask if it formed through a different channel.

Hierarchical mergers---wherein at least one of the components is the product of a binary black hole merger---may occur in dense environments \citep{OLeary:2016ayz,Mapelli:2016vca,Antonini:2016gqe}. 
These systems may be identified by their masses and spins \citep{Fishbach:2017dwv,Gerosa:2017kvu}. 
Being made from smaller black holes, merger remnants are more massive, and their spins are $\sim0.7$ as they are dominated by orbital angular momentum of the merged binary \citep{Buonanno:2007sv}. 
We consider if GW170729's high mass and non-zero spin are evidence for it being a second-generation ($\mathrm{Gen\:2}$) merger. 
Using population distributions from \citet{LIGOScientific:2018jsj} and parameter posterior distributions from \citet{LIGOScientific:2018mvr}, we show that---even under the generous assumption that \emph{all} binary black hole mergers occur in dense clusters---there is not strong evidence that GW170729 is the result of a hierarchical merger.\footnote{The population distributions are available from \href{https://dcc.ligo.org/LIGO-P1800370/public}{dcc.ligo.org/LIGO-P1800370/public} and GW170729's parameter posterior distribution is available from \href{https://dcc.ligo.org/LIGO-P1800324/public}{dcc.ligo.org/LIGO-P1800324/public}.} 

\section{Methodology}

We form an initial ($\mathrm{Gen\:1}$) binary black hole population, drawing masses and spins from the posterior population distributions inferred from gravitational-wave observations \citep{LIGOScientific:2018jsj}.
We use mass Model A---a power-law with a variable exponent, a $5M_{\odot}$ lower cut-off, and a variable upper cut-off---and the non-parametric binned spin model with isotropic alignments.
Model A has been calculated both including and excluding GW170729, allowing exploration of the result's sensitivity to the mass distribution. 
For our default model, we include GW170729 when calculating the probability that it is a first-generation merger (all $10$ binary black holes are first-generation), and exclude it when calculating the probability that it is a second-generation merger (the other $9$ binary black holes are first-generation). 
We use an isotropic distribution of spins under the optimistic assumption that all binary black holes form dynamically in clusters and so may go on to form a new binary.

Remnant spins and masses are calculated following \citet{Healy:2014yta}. 
Second-generation binary black holes are formed from one merger remnant and one initial-population black hole.\footnote{We neglect the probability of forming binary black holes from two merger products.} 
For second-generation mergers we assume that the primary black hole is a merger product, and draw the secondary from the $m_1$ distribution of initial black holes.

\begin{figure}
\begin{center}
\includegraphics[width=0.85\textwidth]{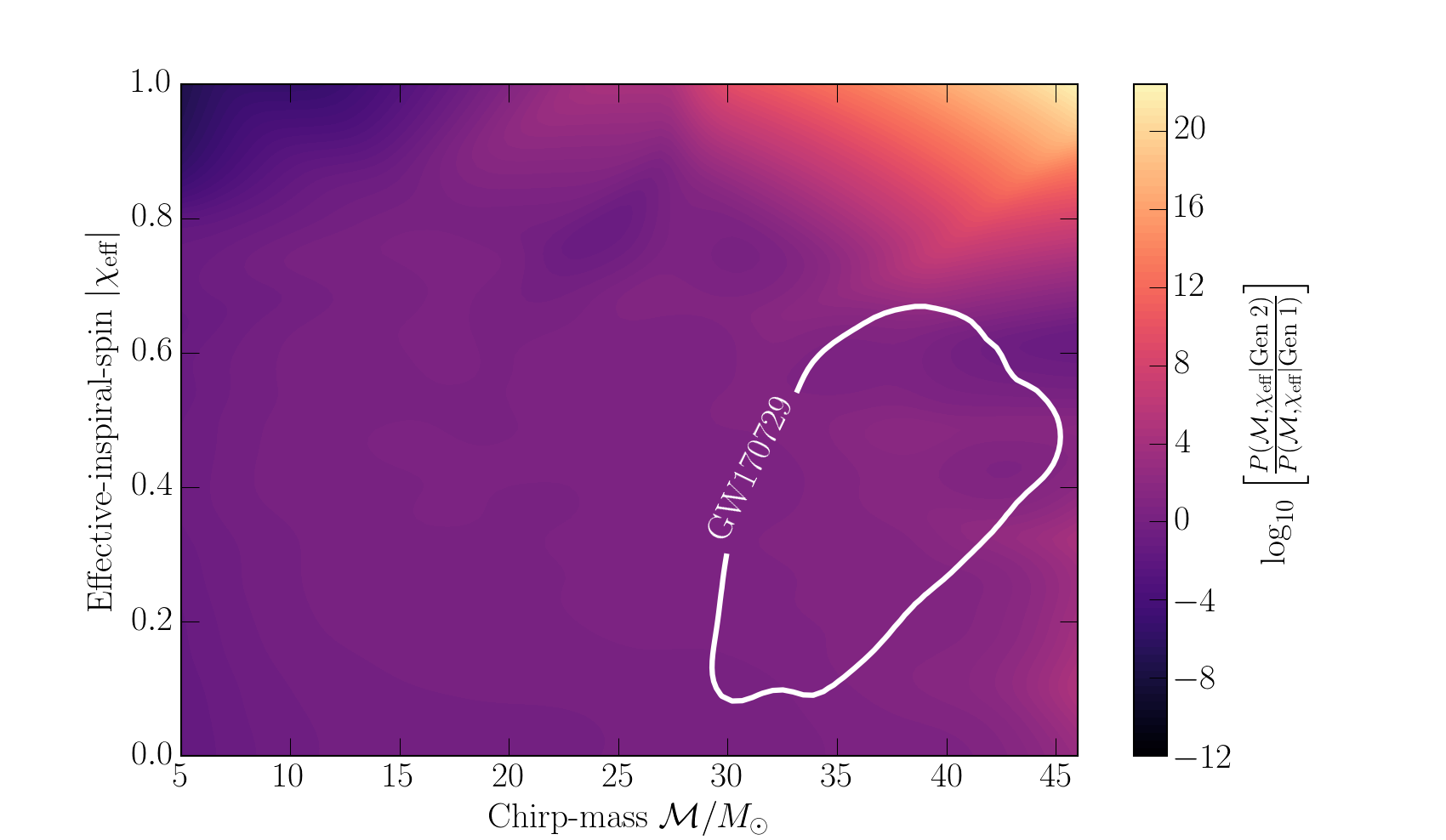}
\caption{Relative likelihood of a binary black hole being second-generation versus first-generation as a function of chirp-mass $\mathcal{M}$ and the magnitude of the effective inspiral spin $|\chi_\mathrm{eff}|$. For comparison, the white contour gives the $90\%$ credible area for GW170729 \citep{LIGOScientific:2018jsj}.}
\label{fig:prob}
\end{center}
\end{figure}

For each generation, we fit a distribution $P(\mathcal{M},\chi_\mathrm{eff}|\mathrm{Gen}\:N)$ over $\mathcal{M}$ and $\chi_\mathrm{eff}$. 
Since we assume isotropic spins for both generations, the probabilities are symmetric about $\chi_\mathrm{eff} = 0$, and we work in terms of $|\chi_\mathrm{eff}|$. 
In Figure~\ref{fig:prob}, we plot the relative probability in favor of $\mathrm{Gen\:2}$ for different points in $\mathcal{M}$--$|\chi_\mathrm{eff}|$ space; higher-mass systems are more likely to be second-generation.
To assess whether GW170729 is a second-generation merger, we calculate the odds ratio 
$P(\mathrm{Gen\:2}|\mathrm{GW170729})/P(\mathrm{Gen\:1}|\mathrm{GW170729})$. The second-generation versus first-generation odds ratio is
\begin{widetext}
\begin{equation}
\frac{P(\mathrm{Gen\:2}|\mathrm{GW170729})}{P(\mathrm{Gen\:1}|\mathrm{GW170729})}=\frac{P(\mathrm{Gen\:2})}{P(\mathrm{Gen\:1})}\left[\frac{\int{}P(\mathrm{GW170729}|\mathcal{M},\chi_\mathrm{eff})P(\mathcal{M},\chi_\mathrm{eff}|\mathrm{Gen\:2})\,\mathrm{d}\mathcal{M}\,\mathrm{d}\chi_\mathrm{eff}}{\int{}P(\mathrm{GW170729}|\mathcal{M},\chi_\mathrm{eff})P(\mathcal{M},\chi_\mathrm{eff}|\mathrm{Gen\:1})\,\mathrm{d}\mathcal{M}\,\mathrm{d}\chi_\mathrm{eff}}\right].
\end{equation}
\end{widetext}
Here, $P(\mathrm{Gen\:N})$ are prior probabilities for each generation, and $P(\mathrm{GW170729}|\mathcal{M},\chi_\mathrm{eff})$ is the likelihood of the observed gravitational-wave signal given the parameters, obtained by dividing the posterior from \citet{LIGOScientific:2018mvr} by the priors used in that analysis. If first- and second-generation
mergers occur at equal rates, then the odds ratio is given by the term in square brackets---the Bayes factor.

\section{Results \& Discussion}

Using the parameters inferred for GW170729 with the \texttt{SEOBNRv3} (\texttt{IMRPhenomPv2}) waveform and our default population model, the estimated Bayes factor is $\sim7$ ($\sim6$).
Both the \texttt{SEOBNRv3} \citep{Pan:2013rra,Taracchini:2013rva} and \texttt{IMRPhenomPv2} \citep{Hannam:2013oca,Khan:2015jqa} waveforms include spin-precession effects, but do not include non-quadrupolar modes; the effect of these (including an independent calculation of the second-generation merger probability) are investigated in \citet{Chatziioannou:2019dsz}.
Calculating both first- and second-generation probabilities using Model A-excluding-GW170729 gives $\sim16$ ($\sim11$), and using Model A-including-GW170729 gives $\sim7$ ($\sim6$).\footnote{
Using other mass models from \citet{LIGOScientific:2018jsj}, \texttt{SEOBNRv3} (\texttt{IMRPhenomPv2}) Bayes factors for Models B and C---which both include GW170729---are $\sim4$ ($\sim3$) and $\sim0.2$ ($\sim0.2$), respectively.
}
Excluding GW170729 gives the highest Bayes factor---the result is sensitive to the upper mass cut-off, and we have selected to exclude the most massive of the observed $10$ binary black holes from the population.
Therefore, it was expected to favor second-generation mergers in this case.
Overall, our results moderately favor GW170729 as a second-generation merger. 
However, adopting a relative prior $P(\mathrm{Gen\:2})/P(\mathrm{Gen\:1}) \lesssim 0.2$ \citep{Rodriguez:2019huv} results in marginally favoring second-generation or favoring a first-generation merger. 
Including the presence of binary black holes merging in the field, which will not undergo multiple mergers, further decreases the probability of a second-generation origin.

In conclusion, we find little evidence that GW170729, despite its mass and spin, is the result of a second-generation merger. 
Results are sensitive to the mass distribution, and in particular the upper mass cut-off; a better understanding of the first-generation population will make it easier to identify second-generation mergers.

\acknowledgments

We thank Chris Pankow for help with the population distributions, and Carl-Johan Haster, Nathan Johnson-McDaniel, and Katerina Chatztziioannou for careful comments on the manuscript. 
This work was partially funded by NSF PHY-1607709 and NSF PHY-1912648. 
CPLB is supported by the CIERA Board of Visitors Research Professorship. 
This work used computing resources at CIERA funded by NSF PHY-1726951.
This document has been assigned LIGO document number \href{https://dcc.ligo.org/LIGO-P1900077/public}{LIGO-P1900077}.

\software{\texttt{matplotlib} \citep{Hunter:2007ouj}
          }

\bibliographystyle{aasjournal}
\bibliography{refs}

\end{document}